\documentclass[12pt]{article}
\usepackage{amsfonts}
\usepackage[cp1251]{inputenc}
\usepackage[english]{babel}
\usepackage{eufrak}
\begin{document}
 \sloppy
\title{Renormalization of Quantum Electrodynamics and Hopf Algebras
}
\author{D.V. Prokhorenko \footnote{Steklov Mathematical Institute, Russian
 Academy of Sciences Gubkin St.8, volovich@mi.ras.ru},
 I.V. Volovich\footnote{Steklov Mathematical Institute, Russian
 Academy of Sciences Gubkin St.8, prokhor@mi.ras.ru}}
\maketitle
\begin{abstract}
In 1999, A. Connes and D. Kreimer have discovered the Hopf algebra
 structure on the Feynman graphs of scalar field theory.
They have found that the renormalization can be interpreted as a
solving of some Riemann --- Hilbert problem. In this work a
generalization of
 their scheme to the case of quantum electrodynamics is
 proposed. The action of the gauge group on the Hopf algebra of
 diagrams are defined and the proof that this action is consistent
 with the Hopf algebra structure is given.

\end{abstract}
\newpage
\section{Introduction}
  \indent The mathematical theory of renormalization (R-operation) was given by
N.N. Bogoliubov and O.S. Parasiuk \cite{1}. K. Hepp has elaborated
on their proofs \cite{2} .\newline
 \indent In 1999, A. Connes and D. Kreimer
\cite{3,4} have introduced the Hopf algebra structure on the
Feynman graphs in scalar field theory with \(\varphi^3 \)
interaction. The Hopf algebras play an important role in the
theory of quantum groups and other noncommutative theories. (About
noncommutative field theory and its relation to p-adic analysis
see \cite{5,6}.) The discovery of the Hopf algebra structure has
clarified the complicated combinatorics of the renormalization
theory. The Feynman amplitudes in this theory belongs to the group
of characters. Let \(U\) be a character corresponding to the
nonrenormalized amplitudes, \(R\) be a character corresponding to
the renormalized amliudes and \(C\) be a character corresponding
to the couterterms. The following equality holds.
\begin{eqnarray}
R=C \star U.
\end{eqnarray}
Here, the star means the group operation in the group of characters.
Let \(U_d\) be a character corresponding to the dimensionally
regularized Feynman amplitudes. \(U_d\) is holomorphic in a certain
punctured neigbourhood of the point \(d=6\) (\(d\) is a parameter of
the dimensional regularization). We can consider \(U_d\) as data of
some Riemann --- Hilbert problem on the group of characters
\cite{7}. A. Connes and D. Kreimer have proven that this problem has
an unique solution, and the positive and negative parts of the
Birkhoff decomposition define renormalized amplitudes and
counterterms respectively. (About future elaboration of this scheme
see \cite{8,9,10}.)\newline \indent In the present paper, we
generalize this scheme to the case of quantum electrodynamics. In
the theory of gauge fields, it is necessary to verify the gauge
invariance of renormalized amplitudes in addition to the proof of
the renormalizability by local counterterms. In the case of quantum
electrodynamics the gauge invariance is expressed in terms of the
Ward identities.\newline \indent So we have the following two
problems.

a) What is the action of the gauge group on the group of characters?

b) Is the action of the gauge group consistent with the Hopf algebra
structure?

 \indent The theory of renormalization of nonabelian
gauge fields was considered in \cite{11}. The extention of the
Hopf algebra formalism to the nonabelian field case requries
additional investigation.
\section{The Hopf algebra of Feynman diagrams}
 The Feynman rules for quantum electrodynamics are described in
\cite{1}. The Feynman graphs of quantum electrodynamics include
the following elements. \newline
 a) The electron lines.\newline
 b) The photon lines.\newline
 c) The vertices of photon-electron interaction. These vertices have precisely
 two external electron lines and one external photon lines.\newline
 d) The vertices of photon self-energy renormalization. These vertices have precisely two
 external photon lines.\newline
 e) The vertices of electron self-energy renormalization. These vertices have precisely two
 external electron lines.\newline
 f) The three-photon vertices,\newline
 g) The four-photon vertices.

 Let us assign to each vertex \(v\) a number
\begin{eqnarray}
\Omega_v=4-3/2n_E-n_F.
\end{eqnarray}
 Here

\(n_E\) is a number of electron lines coming into the vertex
\(v\).

\(n_P\) is a number of photon lines coming into the vertex \(v\).

By definition the space of all polynomials of degree \(\Omega_v\) of
momenta coming into the vertex \(v\) is called the space of vertex
operators at this point and denoted by \(S_v\).

Let \(\Phi\) be a Feynman graph, and \(\varphi\) be a function
which assigns to each vertex \(v\) the element of \(S_v\).

By definition the Feynman diagram is a pair
\(\Gamma=(\Phi,\varphi)\). We will write \(\Phi_\Gamma\),
\(\varphi_\Gamma\) e.t.c. to point out the fact that these
elements corresponds to the diagram \(\Gamma\).

\textbf{Definition.} The algebra \(\mathcal{H}\) of Feynman
diagrams is an unital commutative algebra generated by the pairs
\((\Gamma,\sigma)\). Here \(\Gamma\) is a Feynman diagram and
\(\sigma\) is a tensor-valued distribution with compact support on
the space of momentas of external lines. The tensor structure of
\(\sigma\) corresponds to the types of particles coming into the
diagram. \(\sigma\) is called an external structure.

The generators \((\Gamma,\sigma)\) satisfy the following relations
\begin{eqnarray}
  (\Gamma,\lambda\sigma_1+\mu\sigma_2)=
  \lambda(\Gamma,\sigma_1)+\mu(\Gamma,\sigma_2),\nonumber\\
  \lambda(\Gamma_1,\sigma)+\mu(\Gamma_2,\sigma )=
  (\Gamma_3,\sigma).
  \end{eqnarray}
Here, \(\Gamma_1,\Gamma_2\), and \(\Gamma_3\) are the same as
graphs and for some fixed vertex \(v_0\) of \(\Phi_{\Gamma_1}\)
\begin{eqnarray}
\varphi_{\Gamma_1}(v)=\varphi_{\Gamma_2}(v)=\varphi_{\Gamma_3}(v),\;
\rm if \mit v\neq v_0\nonumber\\
\varphi_{\Gamma_3}(v_0)=\lambda\varphi_{\Gamma_1}(v_0)+\mu\varphi_{\Gamma_2}(v_0).
\end{eqnarray}

  Let us introduce the following notation. Let
  \(\Omega_\Gamma=4-N_F-\frac{3}{2}N_{E}\), where \(
  n_F,n_{E}\) are the numbers of photon and electron lines coming
  into \(\Gamma\), respectively.

  Let \(S_\Gamma\) be a space of all tensor-valued smooth
  functions on the space of momentas of external lines. The tensor
  structure of the elements of \(S_\Gamma\) corresponds to the
  types of particles coming into the diagram. \(S_\Gamma\) is a
  locally convex space with respect to the topology of uniform
  convergence of all derivatives on each compact. Let
  \(S_\Gamma^\star\) be a dual space of the space \(S_\Gamma\).
  \(S_\Gamma^\star\) is a space of all tensor-valued distributions
  with compact support.

  Let  \(S_{\Gamma}^{\Omega_\Gamma}\)
  be a subspace of \(S_{\Gamma}\) of all polynomials of degree
  less or equal to \(\Omega_\Gamma\). Let
   \( {S^{\Omega_\Gamma \ast}_{\Gamma}} \) be a dual space of \(
  S_{\Gamma}^{\Omega_\Gamma}\).
  Let \(\Pi_{\Omega_\Gamma}:S_{\Gamma}\rightarrow
  S_{\Gamma}^{\Omega_\Gamma}\) be a map which assigns to each function \(f\) its Taylor polynomial of degree
  \(\Omega_\Gamma \)
  with the center at zero.
  Let \(B_{\Gamma}^{\Omega_\Gamma}\)=\(\{l_{\Gamma}^{\alpha}\}\),
  \(\alpha \in \mathcal{A}_{\Gamma}\) be some basis of the space
  \( S_{\Gamma}^{\Omega}\).
  Let \( B^{\Omega_\Gamma \ast}_{\Gamma}\) =\(\{l_{\Gamma}^{\alpha
  \ast}\}\) be a dual basis of \(B_{\Gamma}^{\Omega_\Gamma}\).
  To each element  \(l_{\Gamma}^{\alpha \ast} \in B_{\Gamma}^{\Omega_\Gamma \ast} \), assign the
  external structure \(
  \tau_{l_{\Gamma}^{\alpha \ast}}\):=\( l_{\Gamma}^{\alpha\ast}\circ
  \Pi_{\Omega_\Gamma} .\)\newline
  \indent Fix an one-particle irreducible diagram \(\Gamma\).
  Let \(V\) be a set of all vertices of diagram \(\Gamma\). Let \(V'\) be a subset of \(V\).
  Let \(R\) be a set of all lines of diagram \(\Gamma\), \(R_{in}\) be a set of all internal lines of
  \(\Gamma\) and \(R_{ex}\) be a set of all external lines of
  diagram \(\Gamma\). Let \(R'_{in}\) be a subset of all lines from
  \(R_{in}\) such that: \(\forall r \in R'_{in}\) booth of its
  endpoints are the elements of \(V'\). Let \(R'_{ex}\) be a set
  of all pairs \((r,v)\), where \(r\) is a line (external or
  internal) and \(v \in V'\) such that \(r\) has only one endpoint
  from \(V'\) and this endpoint is \(v\).

  Consider the diagram \(\gamma\), such that the set of its
  vertices is \(V'\), the set of its internal lines is
  \(R'_{in}\), the set of its external lines is \(R'_{ex}\), and
  the function \(\varphi_\gamma\) is a restriction of
  \(\varphi_\Gamma\) to \(V'\). If \(\gamma\) is one particle
  irreducible diagram than \(\gamma\) is called an one-particle
  irreducible subdiagm.

  \indent A set of one-particle irreducible nonitersecting subdiagrams of
  \(\Gamma\) is called a subdiagram.
  An element of the subdiagram \(\gamma\) is called a connected component of
  subdiagram. Let \(M\) be a set of all connected components of \(\gamma\). Let \(\alpha\) be a map which to each
element \(i\in M\) assigns an element \(\alpha(i)\) of
\(\mathcal{A}_{\gamma_{i}}\). \(\alpha\)
  is called an multiindex.\newline
  \indent If \(\gamma'\) is one-particle
  irreducible subdiagram, we can replace it by corresponding vertex
 \( v_{0}\). Now, we can essentially  embed
 \(S_{\gamma'}^{\Omega_\gamma'}\) into the space of vertex
 operators at the corresponding vertex \(v_{0}\).
 To each subdiagram  \(\gamma \in \Gamma \) and multiindex
 \(\alpha\), assign an element of the diagram algebra
  \( \gamma_{\alpha} := \prod \limits_{i \in M }(
  \gamma_{i}, \tau_{l_{\gamma_i}^{\alpha(i) \ast}})\). To obtain
  the quotientt diagram \({\Gamma}/{\gamma_{\alpha}}\), we need to replace each connected
  component \(\gamma_{i}\) by the corresponding vertex \(v_{i}\) and
  put \(\varphi(v_{i})=l_{\gamma_{i}}^{\alpha(i)}\). (Here, we use the previous identification.)
  \newline
  \indent \textbf{Definition.} A coproduct \(\Delta\) is
  a homomorphism \(\mathcal{H}\rightarrow \mathcal{H}\otimes \mathcal{H} \)
  defined on generators of \(\mathcal{H}\) by the following
  formula:
  \begin{equation}
  \Delta ((\Gamma,\sigma))=(\Gamma,\sigma)\otimes \mathbf{1}+
  \mathbf{1} \otimes (\Gamma,\sigma) + \sum \limits_{ \emptyset \subset
  \gamma_{\alpha} \subset \Gamma
  } {\gamma_{\alpha}} \otimes
  ( {\Gamma}/{\gamma_{\alpha}},\sigma).
  \end{equation}

  \indent \textbf{ Remark.} In the previous formula, \(\subset\)
  means strong inclusion. The expression \( \gamma_{\alpha} \subset
  \Gamma \) means that \( \gamma \subset
  \Gamma\). The sum is over all nonempty subdiagram and
  and multiindices \(\alpha\).

   Analogusly to \cite{3,4} we have the following theorems.

   \textbf{ Theorem 1.} \textsl{The coproduct \(\Delta\) is well defined
  and does not depend on the special choice of a basis
\(B^{\Omega_\Gamma}_\Gamma\) in the spase
\(S^{\Omega_\Gamma}_\Gamma\).}
  \newline
  \indent \textbf{ Theorem 2.} {The homomorphism \(\Delta\) is coassociative.
  Moreover we can find counit \(\varepsilon\) and antipode
  \(S\) and such that the set \( ({H}, \Delta,
  \varepsilon, S ) \) is a Hopf algebra.  \newline
  \indent A character \(\chi\) on the Hopf algebra \(\mathcal{A}\) is a homomorphism \(\chi:\mathcal{A}\rightarrow\mathbb{C}\).
   The set of all characters is a group with respect the convolution as a group operation.
 \section{ Gauge transformation on the Hopf algebra of diagrams
 of quantum electrodynamics}
 \indent To each one-particle irreducible Feynman diagram, we can assign
 the
 Feynman amplitude \({\Sigma_\Gamma (...,p,...,p',...,k,...)}\). (Here, \(...,p,..,p',..,k,..\) are the momentas, of
 external electron, positron, and photon lines respectively \cite{11,12}.)
 If we regularize the infrared divergences by giving a nonzero mass to
 photon, than \(\Sigma_\Gamma
 (...p...p'..k..)\) is a smooth function of external momentas
 for small nonzero \(z:=d-4\). Here \(d\) is a space-time
 dimensions.
 The Feynman amplitudes define a character on the Hopf algebra by the formula
 \begin{equation}
 U((\Gamma,\sigma))= \int \Sigma_\Gamma (...,p,...,p',...,k,...)
 \sigma (...,p,...,p',...,k,...)...dp...dp'...dk...
 \end{equation}
 \indent \textbf{\(\xi\)-insertion.} \(\xi\)-insertion \cite{1} can
 be made either into a vertex or into an electron line. Let us
 define the
 \(\xi\)-insertion into a vertex. Consider a vertex that has $n$
 entering electron lines, $n$ entering positron lines and $m$ entering photon lines.
 Suppose that the vertex operator at this vertex has the form
 \begin{equation}
 P(p_1,...,p_n|p'_1,...,p'_n|k_1,...,k_m).
 \end{equation}
 Let \(\Omega\) be a degree of
 \(P(p_1,...,p_n|p'_1,...,p'_n|k_1,...,k_m)\).
 By definition \(\xi\)-insertion assigns to this vertex operator
 the vertex operator of degree \(\Omega-1\)
 \begin{equation}
 P_{\xi}^{\mu}(p_1,...,p_n|p'_1,...,p'_n|k_1,...,k_m|k)
 \end{equation}
 defined as
 \begin{eqnarray}
 k^{\mu}P_{\xi}^{\mu}(p_1,...,p_n|p'_1,...,p'_n|k_1,...,k_m|k)=\nonumber \\
 =\sum \limits_{i=1}^{n}\{
 P(p_1,...,p_{i}+k,...,p_n|p'_1,...,p'_n|k_1,...,k_m)-\nonumber\\
 - P(p_1,...,p_n|p'_1,...,p'_{i}+k,...,p'_n|k_1,...,k_m)\}.
 \end{eqnarray}
 After \(\xi\)-insertion the vertex will  have a new external photon appendix. The polynomial defined by this condition
 exists
 but it is defined ambiguously.\newline
 \indent
 To define a \(\xi\)-insertion into the electron line, we need to insert a vertex
 of the photon-electron interaction into the electron line and put
 \(P_\xi^{\mu} (p_1,p'_1|k)=\gamma^\mu\). After this \(\xi\)-insertion the vertex will have
 a new external appendix.\newline
  \indent \textbf{Ward identities.} Let \(\Sigma_\Gamma
 \) be a Feynman amplitude corresponding to the one-particle irreducible diagram
 \(\Gamma\).
 Suppose that the diagram has $n$
 entering electron lines, $n$ entering positron lines, and $m$ entering photon lines.
 Then,
 \begin{eqnarray}
 \sum \limits_{i=1}^{n}
 \{\Sigma_{\Gamma}(p_1,..p_{i}+k,..,p_{n}|p'_{1},...,p'_{n}|k_1,...,k_m)-\nonumber\\
 -\Sigma_{\Gamma}(p_1,...,p_{n}|p'_{1},..p'_{i}+k
 ,..,p'_{n}|k_1,...,k_m)\}= \nonumber \\
 =\sum \limits_{\xi}\{ k^{\mu}
 \Sigma_{\Gamma_{\xi}}^{\mu}(p_1,...,p_{n}|p'_{1},...,p'_{n}|k_1,...,k_m|k)
 \}.
 \end{eqnarray}
In the last line, the sum is over all
 \(\xi\)-insertion into the diagram \(\Gamma\). (The proof is similar
 to the proof in Bogoliubov and Shirkov's monograph \cite{1}.) \newline
 \indent \textbf{Gauge transformation on the Hopf algebra of Feynman diagrams.}
 Let us define now a gauge transformation on the Hopf
 algebra of diagrams \(\mathcal{H}\). For each disribution with compact support \(\alpha \in \mathcal{E'}(\mathbb{R}^4)\)
 define a map \(\delta_\alpha :{H}
 \rightarrow {H} \) by the following formulas:
 \begin{equation}
 \delta_\alpha= \delta'_\alpha+ \delta''_\alpha,
 \end{equation}
 \begin{equation}
 \delta'_\alpha ((\Gamma,\sigma))=(\Gamma, \sigma_\alpha ),\,\,\,\,\delta''_{\alpha}((\Gamma,\sigma))=\sum \limits_{\xi}
 (\Gamma_\xi,\xi_{\alpha}(\sigma)) \label{11}
 \end{equation}
 \begin{eqnarray}
 \sigma_\alpha(p_1,...,p_n|p'_1,...,p'_n|k_1,...,k_m)=\nonumber \\
 =\sum \limits_{i=1}^{n} \int \alpha(k)
 \{\sigma(\widetilde{p}_1,...,\widetilde{p}_{i}-k,...,\widetilde{p}_{n}|
 \widetilde{p}'_{1},...,\widetilde{p}'_{n}|k_1,...,k_m)-\nonumber \\
 -\sigma(\widetilde{p}_1,...,\widetilde{p}_{n}|
 \widetilde{p}'_{1},...,\widetilde{p}'_{i}-k,...,\widetilde{p}_{n}|k_1,...,k_m)\},
 \end{eqnarray}
 where
 \begin{equation}
 \widetilde{p_{i}}=p_{i}+\frac{k}{2n},
 \end{equation}
 \begin{equation}
 \widetilde{p'_{i}}=p'_{i}+\frac{k}{2n}.
 \end{equation}
 The sum in (\ref{11}) is over all \(\xi\)-insertion into diagram
 \(\Gamma\) and
 \begin{eqnarray}
 \xi_{\alpha}(\sigma)(p_{1},...,p_{n}|p'_{1},...,p'_{n}|k_{1},...,k_{m})=\nonumber\\
 =\sigma(\widetilde{p}_{1},...,\widetilde{p}_{n}|\widetilde{p}'_{1},...,\widetilde{p}'_{n}|k_{1},...,k_{m}|k)
 \alpha(k)\otimes k.
 \end{eqnarray}\newline
 Moment \(k\) corresponds to the new external line. \(\sigma\) is a tensor-valued distribution. We use
 the symbol \(\otimes\) in last
  formula in this sense. On the products of elements, we define \( \delta_{\alpha} \) by the
  Leibniz rule \( \delta_{\alpha}( a b )=\delta_{\alpha}(a) b
 +a \delta_{\alpha} (b). \)\newline
 \( \delta_{\alpha} \) is called a gauge transformation on the Hopf
 algebra of diagrams.
 Let $G$ be the group of characters. Then  \(\delta^{\ast}_{\alpha}:G \rightarrow T{G}\), \( U \mapsto
 T_{U}{G}\),
 \begin{equation}
 (\delta^{\ast}_{\alpha}U)(X):=U(\delta_{\alpha}(X)),
 \end{equation}
 \(X \in {H}\). \(\delta^{\ast}\) is called a gauge
 transformation on the group of characters. (Here, \(
 T{G}\) is the tangent bundle of $G$ and \({T_U}{G}\)
 is the tangent space at the point $U$.)
 \newline
 \indent \textbf{Definition.} A character $U$ is called gauge-invariant
 if \(\delta^{\ast}(U)=0. \)\newline
 \indent \textbf{Remark.} Dimensionally regularized Feynman
 amplitudes satisfy the Ward identities; therefore the corresponding characters are gauge
 invariant.
 \newline
 \indent \textbf{Definition.} A character \(U\) is called
 gauge invariant up to degree \(n\) if \(\delta^{\ast}(U)((\Gamma,\sigma))=0
 \) for all diagrams \(\Gamma\) that have at most $n$ vertices.\newline
 \section{Main Results}
 \indent \textbf{Main Theorem.}\textsl{ Let $C$ be a gauge invariant
 character up to degree $n$. Than, for all diagrams \(\Gamma\)
 that have at most $n$ vertices the following relation holds:
 \begin{equation}
 \delta_{\alpha}^{\ast}(C \star U)
 (\Gamma,\sigma)=C\star(\delta_{\alpha}^{\ast} U)(\Gamma,\sigma).
 \end{equation}
 \(\forall \alpha \in \mathbb{E'}(\mathbf{R^4})\)} \newline
 To prove the theorem, we need following lemma. \newline
 \indent \textbf{Lemma.}\textsl{
 Let \(\Gamma\) be a one-particle irreducible diagram, \(\gamma\) be a one-particle
 irreducible subdiagram of \(\Gamma\), and
 \(C\) be a gauge invariant
 character. Then,
 \begin{equation}
 \Gamma':= \sum \limits_{\xi \in \gamma} \sum \limits_{\alpha}
 C((\gamma_{\xi})_{\alpha})({\Gamma_{\xi}}/(\gamma_{\xi})_{\alpha}) =\sum \limits_{\alpha}
  C(\gamma_{\alpha})(\Gamma/\gamma_{\alpha})_{\xi}.
  \end{equation}
 Here, on the left-hand side, \(\xi\)-insertions are made into all vertexes and internal electron
 lines of \(\gamma\), whereas on the right hand side, \(\xi\)-insertions are made into a vertex
 obtained by replacing \(\gamma\) wiht a point.}\newline
  \indent \textbf{Proof} Suppose that there exists a smooth function
  \(f_{C}^{\gamma}\)
 such that
  \(C(\gamma,\sigma)= \int f_{C}^{\gamma}... \sigma ...d p...d k...\)
  Denote by \(v_0\) the vertex of the diagram \( {\Gamma_{\xi}}/{\gamma_{\xi}}\) obtained by
  replacing \(\gamma_{\xi}\) by point. Since \(C\) is gauge
  invariant, we have
  \begin{eqnarray}
  \sum \limits_{i=1}^{n}
  \{
  f^{\gamma}_{C}(p_1,...,p_{i}+k,...,p_{n}|p'_{1},...,p_{n}|k_1,...,k_m)-\nonumber\\
  -f^{\gamma}_{C}(p_1,...,p_{n}|p'_{1},...p_{1}+k...,p_{n}|k_1,...,k_m)\}=\nonumber\\
  =\sum \limits_{\xi} \{
  k^{\mu}( f^{\gamma_{\xi}}_{C})^{\mu}(p_1,...,p_{n}|p'_{1},...,p_{n}|k_1,...,k_m|k)
  \}.
  \end{eqnarray}
  Hence,
  \begin{eqnarray}
  {\Pi}_{\Omega_{\gamma}}
  \{ \sum \limits_{i=1}^{n}
  \{ f^{\gamma}_{C}(p_1,...,p_{i}+k,...,p_{n}|p'_{1},...,p_{n}|k_1,...,k_m)-\nonumber\\
  -f^{\gamma}_{C}(p_1,...,p_{n}|p'_{1},...,p'_{i}+k,...,p'_{n}|k_1,...,k_m) \}\}=\nonumber\\
  = {\Pi}_{\Omega_{\gamma}} \{\sum \limits_{\xi} \{
  k^{\mu}( f^{\gamma_{\xi}}_{C})^{\mu}(p_1,...,p_{n}|p'_{1},...,p_{n}|k_1,...,k_m|k)
  \}\}=\nonumber\\
  =k^{\mu} {\Pi}_{\Omega_{\gamma_\xi}} \{\sum \limits_{\xi}\{
  ( f^{\gamma_{\xi}}_{C})^{\mu}(p_1,...,p_{n}|p'_{1},...,p_{n}|k_1,...,k_m|k)
  \}\}.
  \end{eqnarray}
  But the last term in this formula is equal to \(
  (\varphi_{\Gamma'}(v_{0}))^{\mu} k^{\mu} \). Therefore,
  \((\varphi_{\Gamma'}(v_{0}))^{\mu} \) is equal to the result of
  the \(\xi\)-insertion into \(v_{0}\).  In order to prove this lemma in the general case, it is enough to
  note, that there exist smooth function
  \(f_{C}^{\gamma}\) such that \newline
  \(C((\gamma,{\Pi}_{\Omega}^{\ast}(\sigma)))= \int
  f_{C}^{\gamma}...
  {\Pi}_{\Omega}^{\ast}( \sigma) ...d p...d k...\).\newline
  The lemma is proved. \newline
\indent \textbf{Proof of the main theorem.}  We have
  \begin{eqnarray}
  \delta_{\alpha}^{\ast}(C \star U)((\Gamma,\sigma))=(C \star
  U)\{\delta_{\alpha}((\Gamma,\sigma))\}=\nonumber\\
  =(C\star
  U)\{\delta'_{\alpha}((\Gamma,\sigma))+\delta''_{\alpha}((\Gamma,\sigma))\},
  \end{eqnarray}
  \begin{eqnarray}
  (C\star U) (\delta'_{\alpha}((\Gamma,\sigma)))=(C\star U)((\Gamma,\sigma_{\alpha}))=\nonumber\\
  =C((\Gamma,\sigma_{\alpha}))+ \sum \limits_{\emptyset \subseteq
  \gamma_{\beta} \subset \Gamma } C(\gamma_{\beta})U(({\Gamma
  /{\gamma_\beta}},\sigma_{\alpha})), \label{3}
  \end{eqnarray}
  \begin{eqnarray}
  (C \star U)(\delta''_{\alpha}((\Gamma,\sigma)))=\sum \limits_{\xi}
  (C\star U)(( \Gamma_{\xi},\xi_{\alpha}(\sigma)))=\nonumber\\
  =({\delta''_{\alpha}}^{\ast} C)((\Gamma,\sigma))+
  ({\delta''_{\alpha}}^{\ast} U)((\Gamma,\sigma))+\sum \limits_{\xi}
  \sum \limits_{\emptyset \subset \gamma'_{\beta} \subset
  \Gamma_{\xi}} C(\gamma'_{\beta}) U(({\Gamma_{\xi}/\gamma'_{\beta}
  }, \xi_{\alpha}(\sigma))). \label{24}
  \end{eqnarray}
  To each subdiagram \(\gamma\) of \(\Gamma\) and
  \(\xi\)-insertion into \(\Gamma\), assign a
  subdiagram \(\gamma'\) of a diagram \(\Gamma_{\xi}\) in the
  following way: if the \(\xi\)-insertion is not a \(\xi\)-insertion into
  \(\gamma\), then
  \(\gamma'=\gamma\). If the \(\xi\)-insertion is a \(\xi\)-insertion into \(\gamma\), than
  \(\gamma'=\gamma_{\xi}\).
  The subdiagram \(\gamma'\subseteq \Gamma_{\xi}\) is proper, if and only if
  \(\gamma\) is a proper subdiagram of \(\Gamma\). We have the following expression
  for the last term in (\ref{24}):
  \begin{eqnarray}
  \sum \limits_{\emptyset \subset \gamma_{\beta} \subset \Gamma }
  \sum \limits_{\xi \in \gamma } C( \gamma'_{\beta}) U(({{\Gamma_\xi}
  /
  (\gamma_{\beta})_{\xi}}, \xi_{\alpha}(\sigma)))+\nonumber\\
  +\sum \limits_{\emptyset \subset \gamma_{\beta} \subset \Gamma }
  \sum \limits_{\xi \notin \gamma} C( \gamma_{\beta}) U((({\Gamma /
  \gamma_{\beta}})_{\xi}, \xi_{\alpha}(\sigma))).
  \end{eqnarray}
  The first term of this expression is equal to
  \begin{equation}
  \sum \limits_{\emptyset \subset \gamma_{\beta} \subset \Gamma }
  {\sum \limits_{\xi }}' C( \gamma_{\beta}) U((({\Gamma /
  \gamma_{\beta}})_{\xi}, \xi_{\alpha}(\sigma))).
  \end{equation}
  Here, the sum is over all \(\xi\)-insertion into the vertices
  of the
  diagram
  \( \Gamma / \gamma_{\beta}\) obtained by replacing all
  connected components of \(\gamma\) by points. We have
  \begin{eqnarray}
  ({\delta''_{\alpha}}^{\ast} (C\star U))((\Gamma,\sigma))
  =({\delta''_{\alpha}}^{\ast}
  C)((\Gamma,\sigma))+({\delta''_{\alpha}}^{\ast}
  U)((\Gamma,\sigma))+ \nonumber \\
  +\sum \limits_{\emptyset \subset \gamma_{\beta} \subset \Gamma }
  \sum \limits_{\xi \in {\Gamma/\gamma}} C( \gamma_{\beta})
  U((({\Gamma / \gamma_{\beta}})_{\xi},
  \xi_{\alpha}(\sigma)))=\nonumber\\
  =({\delta''_{\alpha}}^{\ast}
  C)((\Gamma,\sigma))+({\delta''_{\alpha}}^{\ast} U)((\Gamma,\sigma))+
  \sum \limits_{\emptyset \subset \gamma_{\beta} \subset \Gamma } C(
  \gamma_{\beta})({\delta''_{\alpha}}^{\ast} U)(({\Gamma /
  \gamma_{\beta}},\sigma)). \label{30}
  \end{eqnarray}
  It follows from (\ref{3}) and (\ref{30}) that
 \begin{equation}
 \delta_{\alpha}^{\ast} (C \star U)((\Gamma,\sigma))=C \star
 (\delta_{\alpha}^{\ast}U)((\Gamma,\sigma)).
 \end{equation}
  The theorem is completely proved.\newline
   \indent \textbf{Remark.} Actually, we have proved the following proposition.
  if
 \(C\) is gauge invariant character up to degree \(n-1\) and \(U\) is a character, than
 \begin{eqnarray}
 \delta_{\alpha}^{\ast} \{ U (\bullet )
 +  \sum \limits_{\emptyset
 \subset \gamma_{\alpha} \subset \bullet } C( \gamma_{\alpha})
 U( \bullet / {\gamma_{\alpha}}) \}=\nonumber \\
 =\delta_{\alpha}^{\ast} U (\bullet )+ \sum \limits_{\emptyset
 \subset \gamma_{\alpha} \subset \bullet } C( \gamma_{\alpha})
 ( \delta_{\alpha}^{\ast}U) (\bullet /{\gamma_{\alpha}}).
 \end{eqnarray}
 The following theorem holds. \newline
 \indent \textbf{Theorem 3.} \textsl{The set of all gauge-invariant characters is a subgroup
 of the group of characters.}\newline
 \indent\textbf{Proof.} The fact that a product of two gauge
 invariant characters is a gauge invariant character follows
 from the main theorem. The counit is obviously gauge invariant.
 We can prove the gauge invariance of \(U^{-1}\) using the remark
 to the main theorem and the following relation:
 \begin{equation}
 U^{-1}(\Gamma)=-U(\Gamma)- \sum \limits_{\emptyset \subset
 \gamma_{\alpha} \subset \Gamma} U^{-1}(\gamma_{\alpha}) U(\Gamma /
 \gamma_{\alpha}).
 \end{equation}
 \newline
 \indent \textbf{The Riemann---Hilbert problem.}
 Let \(U_{z}\) be a character holomorphic in \(z\) in a small
 punctured neighbourhood  \(O \setminus \{0\}\) of zero. To solve the Rieman-Hilbert
 problem,
 we must to find the Birkhoff decomposition of \(U_{z}\), i.e., to find
 characters \(R_{z}\) and \(C_{z}\) such that \(R_{z}\)
 holomorphic in \(\setminus \{0\}\), \(C_{z}\) holomorphic in \(\mathbf{C}\setminus
 \{0\}\), the following gluing conditions holds
 \begin{eqnarray}
 R_{z}=C_{z} \star U_{z}
 \end{eqnarray}
 in \(O\setminus \{0\} \),
 and \(C_{z} \rightarrow \varepsilon\) as \(z \rightarrow \infty\).
 This problem has an unique solution. This fact follows from the Liouville theorem. \newline
 \indent \textbf{Remark.} We say that the character \(U_{z}\) is continuous, holomorphic
 ect. wiht in \(z\) if \(U_{z}(X)\) is continuous, holomorphic
 ect. wiht in \(z\) for all \(X \in H \). \newline
 \indent  The following theorem generalize the result of Connes and
 Kreimer for the scalar model to the case of quantum
 electrodynamics.\newline
 \indent  \textbf{Theorem 4.} \textsl{A solution of the Riemann-Hilbert problem
 exists.}
 \newline
 \indent
 \textbf{Proof.} We follow the Connes-Kreimer proof for the
 scalar field. A solution of the problem is given by the explicit
 formulas
 \begin{equation} C_{z}(\Gamma)=-{T}(U_{z}(\Gamma)+ \sum
 \limits_{\emptyset \subset \gamma_{\alpha} \subset \Gamma}
 C_{z}(\gamma_{\alpha}) U_{z}(\Gamma / \gamma_{\alpha})),
 \end{equation}
 \begin{equation}
 R_{z}(\Gamma)=(1-{T})(U_{z}(\Gamma)+ \sum
 \limits_{\emptyset \subset \gamma_{\alpha} \subset \Gamma}
 C_{z}(\gamma_{\alpha}) U_{z}(\Gamma / \gamma_{\alpha})).
 \end{equation}
 Here by definition the operator \({T}\) assigns to each function
 \(f\) holomorphic in \( O\setminus \{0\}\) its pole part.\newline
 \indent \textbf{Remark.} Let \(U_{z}\) be a character
 corresponding to the dimensionally regularized Feynman
 amplitudes.In this case
 \(C_{z}\) corresponds to counterterms, and \(R_{z}\) corresponds to the
 set of renormalized amplitude.\newline
 \indent \textbf{Theorem 5.}\textsl{ If the data of the Riemann-Hilbert problem is gauge-invariant,
 than the solution of this problem is gauge-invariant too.} \newline
 \indent \textbf{Proof.} This fact follows from the remark to the main theorem and
  the obvious fact that the gauge transformation commutes with \(T\).\newline
 \section{Conclusion}
 In this paper, we have given a generalization of Connes-Kreimer
 method in renormalization theory to the case of quantum electrodynamics.
 We have introduced the Hopf algebra of diagrams,
 that generalize the corresponding construction of Connes and
 Kreimer. We have obtained two main results. The first one is that
 the set of gauge-invariant characters is a group. The second one
 is that the Riemann ---  Hilbert problem has a gauge-invariant solution
 if the data of this problem is gauge-invariant. It is
 interesting to generalize these results to the case of nonabelian
 gauge fields.
 \section{Acknowledgments}
 The first author (D.V.P.) would like to thank M.G. Ivanov for very
 interesting discussions.

 This work was partially supported by the Russian
 Foundation of Basis Reasearch (project 05-01-008884), the grand
 of the president of the Russian Federation (project
 NSh-1542.2003.1) and the program "Modern problems of
 theoretical mathematics" of the mathematical Sciences department
 of the Russian Academy of Sciences.

 \end{document}